\documentclass[9pt]{article}
\usepackage[left=15 mm,right=10 mm,
top=20 mm,bottom=20 mm]{geometry}
\usepackage{graphicx}
\usepackage{bm}
\usepackage{commath}
\usepackage{xcolor}
\usepackage{soul}
\usepackage{lipsum}
\usepackage{mathtools}
\usepackage{cuted}
\usepackage{amsmath}
\usepackage{float}
\usepackage[normalem]{ulem}
\usepackage{braket}
\usepackage{url}
\usepackage{caption}
\providecommand{\keywords}[1]
{
	\small	
	\textbf{\textit{Keywords---}} #1
}

\begin{document}

\title{Fluorescence of rubidium vapor in a transient interaction regime}

\author{Artur Aleksanyan*, Svetlana Shmavonyan, Emil Gazazyan, Aleksandr Khanbekyan,\\Hrayr Azizbekyan, Marina Movsisyan, Aram Papoyan \\\\ \small Institute for Physical Research, NAS of Armenia, Ashtarak-2, 0203 Armenia \\\\ \small Corresponding author: arthuraleksan@gmail.com}

\maketitle

\begin{abstract}
We have studied modification of the fluorescence spectra of a room-temperature atomic rubidium vapor in the region of $^{85}$Rb and $^{87}$Rb D$_2$ line while changing the temporal rate of linear (triangular) scanning of laser radiation frequency. Increase of the ramping speed over certain value ($\approx$ 10$^6$ MHz/s) results in essential modification of magnitudes of individual atomic transitions, different on rising and falling slopes, which characterize transition from a steady-state interaction regime to a transient one. Our experimental results are well consistent with the developed theoretical model. The obtained results can be used for determination of atomic system parameters such as ground-state relaxation rate. Possible follow-up actions on addressed control of atomic levels population is discussed.
\end{abstract}


\section{Introduction}
\label{sec:Sec.1}

Resonant interaction of narrow-linewidth cw laser radiation with atomic vapors of alkali metals is intensely studied in the past decades, driven by fundamental interest and emerging important applications. Most of these studies deal with a steady-state regime of interaction of atomic ensemble with resonant light required for establishment of the relevant processes.

To the best of our knowledge, there are just a few works on atomic spectroscopy with cw excitation radiation, where transient resonant interaction processes were studied. Particularly, theoretical and experimental investigations of processes under dynamic excitation of atomic media with modulated cw laser radiation were done for nonlinear magneto-optical processes \cite{Alexandrov_2005}, saturation spectroscopy \cite{Thornton_2011}, four-wave mixing \cite{Becerra_2010}, coherent population trapping  \cite{Khripunov_2016,Yudin_2017}. Analytical solutions of temporal evolution of populations in optically-pumped atoms were obtained in \cite{Noh_2016}. Besides transient processes imposed by temporally-modulated laser radiation, dynamic effects were studied also for spatial Ramsey schemes, such as dark Raman resonances caused by interference \cite{Grujic_2008}.

When laser radiation frequency is tuned to an atomic transition, the atom can undergo many cycles of absorption and emission, which eventually lead to establishment of the steady-state atomic response. There are two main factors, which determine interaction time of an individual atom with the laser field in conventional atomic spectroscopy experiment. Firstly, the interaction time can be limited by a time of flight of an atom through a laser beam. For the room-temperature alkali vapor, the mean atomic velocity is $\sim$200 m/s, and for the 1 mm-diameter laser beam, the time of flight of an atom traversing the laser beam at 90$^{\circ}$ angle is $\tau_{tof}$ = 5 $\mu$s. Secondly, the interaction time can be determined by the temporal rate of linear scanning of laser radiation frequency employed in many experiments. For example, when scanning $\Delta \omega_L$ = 2$\pi \times$10 GHz frequency interval around the resonance line by applying triangular modulation pulses with repetition frequency of $f_s$ = 50 Hz, the interaction time of an individual atom with $\gamma_{nat}$ = 2$\pi \times$6.07 MHz natural linewidth will be

\begin{equation}
\label{eq::intscan}
\tau_s=\frac{\gamma_{nat}}{2f_s\Delta\omega_L}\approx 6\: \mu s.
\end{equation}

In most of spectroscopic experiments, the scanning is slow enough, so that the interaction time is restricted by a flight time.

Besides the above mentioned physical limitations, the resonant interaction of atom with laser field is governed by the laser electric field amplitude \textbf{E$_0$} and atomic transition dipole moment \textbf{d$_{i,j}$}, characterized by a Rabi frequency $\Omega_{i,j}$ = \textbf{d$_{i,j}$}\textbf{E$_0$}$/\hbar$. The corresponding experimental (measurable) parameter is laser radiation intensity $I_L$:

\begin{equation}
	\label{eq::intensity}
	I_L=\frac{1}{2}\epsilon_0nc\lvert{E_0}\rvert^2=\frac{\epsilon_0nc\hbar^2\Omega^2_{i,j}}{d^2_{i,j}},
\end{equation}

\noindent where $\epsilon_0$ is the vacuum permittivity, and $n$ is the refractive index. Taking into account possible detuning $\Delta$ of the laser radiation frequency from the atomic transition (including also the spectral linewidth of laser radiation when it exceeds $\gamma_{nat}$), generalized (effective) Rabi frequency should be considered:

\begin{equation}
\label{eq::effectrabi}
\widetilde{\Omega}_{i,j}=\sqrt{\Omega^2_{i,j}+\Delta^2}.
\end{equation}

Study of transient effects in resonant interaction of cw laser radiation with atomic media is of practical interest for two reasons. First, as is seen from the above mentioned, it can be used for determination of experimental parameters such as relaxation rates. Second, dynamic transient processes can be utilized for heralded control of population in atomic systems, resembling excitation by $\pi$ and $\pi/2$ pulses in a Rabi cycle. Studies of transient processes involving fluorescence spectra are of particular interest, since the fluorescence can serve as a direct measure of population of excited atomic states. 

In this article, we present the results of theoretical and experimental studies of Rb D$_2$ line fluorescence spectra while changing the rate of linear scanning of laser radiation frequency by four orders of magnitude. Our primary aim was to determine the most appropriate temporal conditions for efficient atomic population control. Also, we were aimed at determination of important relaxation parameters of atomic system based on the fitting of experimental results by our theoretical model.

\begin{figure}[ht!]
	\begin{center}
	\includegraphics[width=240pt]{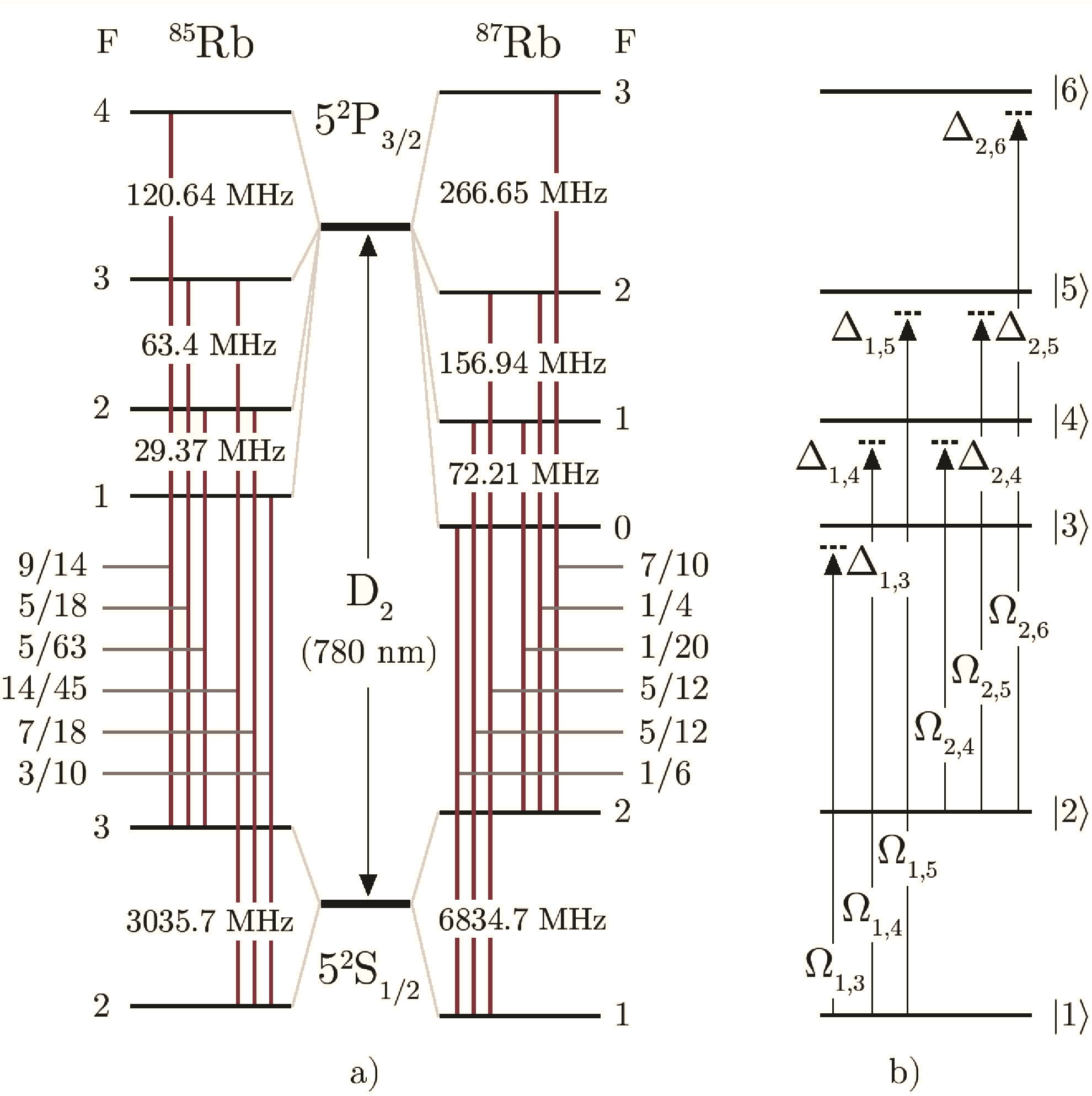}
	\caption{\label{fig:Fig.1} a) Hyperfine structure of rubidium D$_2$ line and individual optical transitions for $^{85}$Rb \cite{SteckRb85} and $^{87}$Rb \cite{SteckRb87} with indicated relative strengths. b) Scheme of the theoretical model with notations of the parameters.}
	\end{center}
\end{figure}

\section{Theoretical model}
\label{sec:Sec.2}

We employ a density matrix model \cite{blum2012density} to simulate the resonant fluorescence on the hyperfine transitions $^{85}$Rb F$_g$=2,3 $\rightarrow$ F$_e$=1,2,3,4 and $^{87}$Rb F$_g$=1,2 $\rightarrow$ F$_e$=0,1,2,3 of atomic D$_2$ line (see Fig.\ref{fig:Fig.1}a), developed upon excitation of the atomic system by laser radiation with a frequency scanned across the hyperfine transitions manifold (see the system diagram in Fig.\ref{fig:Fig.1}b). 

For this system, the time-dependent Liouville -- von Neumann equation will read:

\begin{equation}
\label{eq:DME}
i\hbar\dfrac{\partial \rho}{\partial t}=[H,\rho]-R(\rho),
\end{equation}

\noindent where $\rho$ is 6$\times$6 dimentional density matrix with diagonal elements $\rho_{i,i}(t)$ representing the population of $\ket{i}$-th state, and off-diagonal elements $\rho_{i,j}(t)$ representing coherences linked with $\ket{i} \rightarrow \ket{j}$ transitions, $H$ is the Hamiltonian of the system, and $R(\rho)$ is the relaxation matrix. As we deal with non-stationary (transient) interaction regime caused by fast frequency scanning, we consider a time dependent problem. 
The initial condition for \eqref{eq:DME}
is $\rho_{1,1}(0)+\rho_{2,2}(0)=1$. Taking into account the magnetic sublevels manifold, the initial ground state populations are $\rho_{1,1}(0)=5/12$, $\rho_{2,2}(0)=7/12$ for $^{85}$Rb, and  $\rho_{1,1}(0)=3/8$, $\rho_{2,2}(0)=5/8$ for $^{87}$Rb \cite{SteckRb85,SteckRb87}. The time dependent Hamiltonian will have the following form:

\begin{equation}
\label{eq:Hamiltonian}
H(t)=\left( 
\begin{matrix}
0 & 0 & \Omega_{1,3} e^{-i\Delta_{1,3}(t)t} & \Omega_{1,4} e^{-i\Delta_{1,4}(t)t} & \Omega_{1,5} e^{-i\Delta_{1,5}(t)t} & 0\\
0 & 0 & 0 & \Omega_{2,4} e^{-i\Delta_{2,4}(t)t} & \Omega_{2,5} e^{-i\Delta_{2,5}(t)t} & \Omega_{2,6} e^{-i\Delta_{2,6}(t)t} \\
\Omega_{1,3} e^{i\Delta_{1,3}(t)t} & 0 & 0 & 0 & 0 & 0 \\
\Omega_{1,4} e^{i\Delta_{1,4}(t)t} & \Omega_{2,4} e^{i\Delta_{2,4}(t)t} & 0 & 0 & 0 & 0 \\
\Omega_{1,5} e^{i\Delta_{1,5}(t)t} & \Omega_{2,5} e^{i\Delta_{2,5}(t)t} & 0 & 0 & 0 & 0 \\
0 & \Omega_{2,6} e^{i\Delta_{2,6}(t)t} & 0 & 0 & 0 & 0\\
\end{matrix} 
\right),
\end{equation}

\noindent where $\Omega_{i,j}=\dfrac{\bm{{E}_{0}}\bm{{d}_{i,j}}}{\hbar}$ are the matrix elements of the Rabi frequency with $\bm{{d}_{i,j}}$ the matrix elements of the dipole moment for the respective transitions \cite{SteckRb85,SteckRb87}, $\bm{E_0}$ is the amplitude of classical electric field interacting with atomic medium of $^{87}$Rb and $^{85}$Rb, and $\Delta_{i,j}(t)$ are one-photon detunings of the scanning laser field from atomic resonances. For periodic triangular temporal modulation of the laser radiation frequency, we can write

\begin{equation}
\Delta_{i,j}(t)=\Delta_{i,j}^0+\dfrac{\Delta}{\pi}\arcsin{(\cos{2\pi f_s t})},
\end{equation}

\noindent where $\Delta$ the spectral range of scanning, $f_{s}$ is the triangular modulation frequency, $i=1,2$, and $j=3,4,5,6$, see Fig.\ref{fig:Fig.1}b.
Employing this modulation, the radiation frequency will linearly increase / decrease in time, so that the laser field will be consecutively in resonance with all the groups of transitions: $^{87}$Rb $F_g=2 \rightarrow F_e=1,2,3$, $^{85}$Rb $F_g=3 \rightarrow F_e=2,3,4$, $^{85}$Rb $F_g=2 \rightarrow F_e=1,2,3$, and $^{87}$Rb $F_g=1 \rightarrow F_e=0,1,2$, in direct (rising frequency) and reverse (falling frequency) order. 

The relaxation matrix R($\rho$) involves all the relaxation processes in the system:

\begin{equation}
\label{eq:Lindblad}
R(\rho)=\left(
\begin{matrix}
\Gamma^{'}&(\gamma_0+\gamma_{tot})\rho_{1,2}&\gamma_{tot}\rho_{1,3}&\gamma_{tot}\rho_{1,4}&\gamma_{tot}\rho_{1,5}&\gamma_{tot}\rho_{1,6}\\
(\gamma_0+\gamma_{tot})\rho_{2,1}&\Gamma^{''}&\gamma_{tot}\rho_{2,3}&\gamma_{tot}\rho_{2,4}&\gamma_{tot}\rho_{2,5}&\gamma_{tot}\rho_{2,6}\\
\gamma_{tot}\rho_{3,1}&\gamma_{tot}\rho_{3,2}&(\Gamma_{3,1}+\gamma_0)\rho_{3,3}&\gamma_{tot}\rho_{3,4}&\gamma_{tot}\rho_{3,5}&\gamma_{tot}\rho_{3,6}\\
\gamma_{tot}\rho_{4,1}&\gamma_{tot}\rho_{4,2}&\gamma_{tot}\rho_{4,3}&(\Gamma_{4,1}+\Gamma_{4,2}+\gamma_0)\rho_{4,4}&\gamma_{tot}\rho_{4,5} & \gamma_{tot}\rho_{4,6}\\
\gamma_{tot}\rho_{5,1}&\gamma_{tot}\rho_{5,2}&\gamma_{tot}\rho_{5,3}&\gamma_{tot}\rho_{5,4}&(\Gamma_{5,1}+\Gamma_{5,2}+\gamma_0)\rho_{5,5}&\gamma_{tot}\rho_{5,6}\\
\gamma_{tot}\rho_{6,1}&\gamma_{tot}\rho_{6,2}&\gamma_{tot}\rho_{6,3}&\gamma_{tot}\rho_{6,4}&\gamma_{tot}\rho_{6,5}&(\Gamma_{6,2}+\gamma_0)\rho_{6,6}\\
\end{matrix}\right)
\end{equation}

\noindent where the following notations are used:

\begin{align*}
\Gamma^{'}=\gamma_0(\rho_{1,1}-\rho^0_{1,1})-\Gamma_{3,1}\rho_{3,3}-\Gamma_{4,1}\rho_{4,4}-\Gamma_{5,1}\rho_{5,5},& \\
\Gamma^{''}=\gamma_0(\rho_{2,2}-\rho^0_{2,2})-\Gamma_{4,2}\rho_{4,4}-\Gamma_{5,2}\rho_{5,5}-\Gamma_{6,2}\rho_{6,6}.
\end{align*}

\noindent 
Here $\Gamma_{i,j}$ is the natural decay rate of the corresponding excited state; $\gamma_0$ is the relaxation rate of the lower energy levels to the equilibrium isotropic state \cite{Yudin_2017}; $\gamma_{tot}$ is the total broadening rate comprising radiative damping, collisional broadening, laser radiation linewidth, and inhomogeneous (Doppler) broadening making a dominant contribution ($\gamma_{tot} \approx \gamma_{Dop}$). All the rate values used in theoretical calculations have been normalized to the natural decay rate for the Rb D$_2$ line: $\gamma_{nat} = 2\pi \times$6.07 MHz. 

The time-dependent fluorescence spectra are then calculated numerically using the following formula: 

\begin{equation}
\Phi_{t}(t)=\sum_{i=3,4,5,6} \Gamma_{i,j}\rho_{i,i}(t).
\end{equation}

\section{Experiment and numerical simulation}
\label{sec:Sec.3}

\begin{figure}[hb!]
	\begin{center}
		\includegraphics[width=135pt]{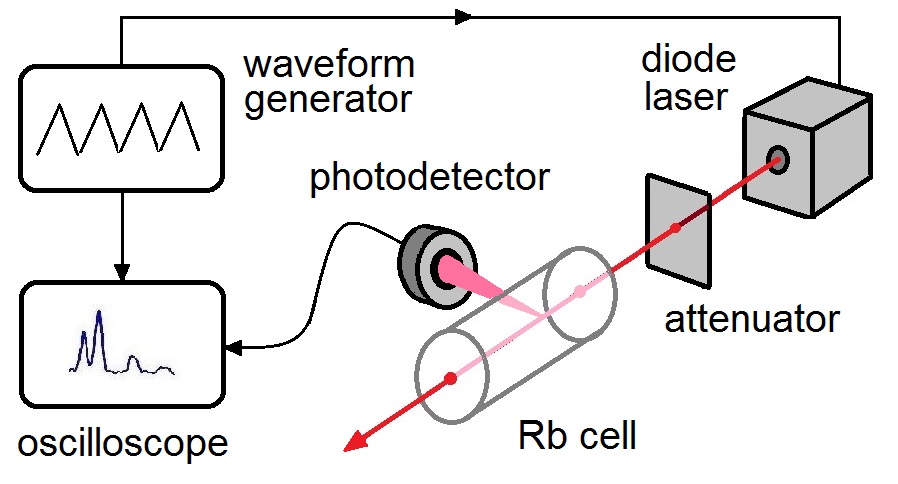}
		\caption{\label{fig:Fig.2} Schematic drawing of the experimental setup.}
	\end{center}
\end{figure}

Experimental measurements were done on a simple setup schematically depicted in Fig.\ref{fig:Fig.2}. Collimated linearly-polarized radiation from a free-running single-frequency laser diode (maximum power 25 mW; spectral linewidth 15 MHz) with 2 mm beam diameter was directed into a glass cell (135 mm-long, 20 mm-diameter, no antirelaxation coating, no added buffer gas) with a side arm containing natural rubidium. The choice of a free-running laser was conditioned by a necessity of a fast linear frequency scanning which, unlike external PZT-driven cavity diode lasers, is easily realizable by modulation of an injection current. The cell was kept at a room temperature (22 $^\circ$C), which corresponds to number density of rubidium atoms $N_{Rb}$ = 5$\times$10$^9$ cm$^{-3}$. A fast linear photodetector was placed at 90$^\circ$ to the laser beam propagation direction, closer to the entrance window. 

In order to scan the laser radiation frequency across the spectral region of atomic D$_2$ line, covering Doppler-overlapped hyperfine transition groups $^{87}$Rb F$_g$=2 $\rightarrow$ F$_e$=1,2,3, $^{85}$Rb F$_g$=3 $\rightarrow$ F$_e$=2,3,4, $^{85}$Rb F$_g$=2 $\rightarrow$ F$_e$=1,2,3, and $^{87}$Rb F$_g$=1 $\rightarrow$ F$_e$=0,1,2 (typically 11 GHz range), the laser diode injection current was modulated by periodic triangular pulses from Siglent SDG5082 waveform generator. The scanning rate and frequency range were controlled by changing the generator frequency and amplitude, respectively. It was possible to fine tune the laser radiation frequency by applying a bias (offset) to the generator signal. Fluorescence signal from the photodetector (photodiode with operational amplifier) was recorded by a digital storage oscilloscope Tektronix TDS-3032B. The maximum used scanning frequency was limited by the temporal response of the photodetector ($\tau_{det}\approx$ 5 $\mu$s).

\begin{table}[h]
	\footnotesize
	\caption{Linkage between the temporal parameters of the experiment}
	\centering
	\renewcommand{\arraystretch}{0.9}
	\begin{tabular}{c rrr} 
		\multicolumn{1}{c}{No.}&$f_s$ (Hz)&$\tau_{\pm}$ (ms)&$\frac{S}{2\pi}$ (MHz/$\mu$s) \\
		\hline\\
		1&1&500&0.02224\\
		2&2.5&200&0.0556\\
		3&5&100&0.1112\\
		4&10&50&0.2224\\
		5&25&20&0.556\\
		6&50&10&1.112\\
		7&100&5&2.224\\
		8&250&2&5.56\\
		9&500&1&11.12\\
		10&1000&0.5&22.24\\
		11&2500&0.2&55.6\\
		12&5000&0.1&111.2\\
		13&10000&0.05&222.4\\
		\hline 
	\end{tabular}
	\label{tab:conversion}
\end{table}

The scanning time (period) itself can not be considered as a physical parameter, since the resonant interaction time of an individual atom with laser radiation depends also on the spectral range covered by scanning. A real physical meaning should be attributed to the scanning rate defined as $S=\partial\omega/\partial t$. In addition, the interaction time is also affected by the homogeneous broadening width and laser radiation linewidth. For this reason, to facilitate interpretation of the results, only two experimental parameters were varied throughout our measurements: laser radiation power $P_L$ and triangular modulation frequency $f_s$. The spectral range $\Delta \omega_s$ = 2$\pi \times$11.12 GHz was kept invariable both on descending ($\omega_-$) and ascending ($\omega_+$) wings. Moreover, also the frequency positions of hyperfine transitions in the scanning spectral range were kept unchanged, independently of $f_s$ value. In these conditions the scanning rate can be determined by simple rescaling of modulation (scanning) frequency:

\begin{figure}[h!]
	\begin{center}
	\includegraphics[width=323pt]{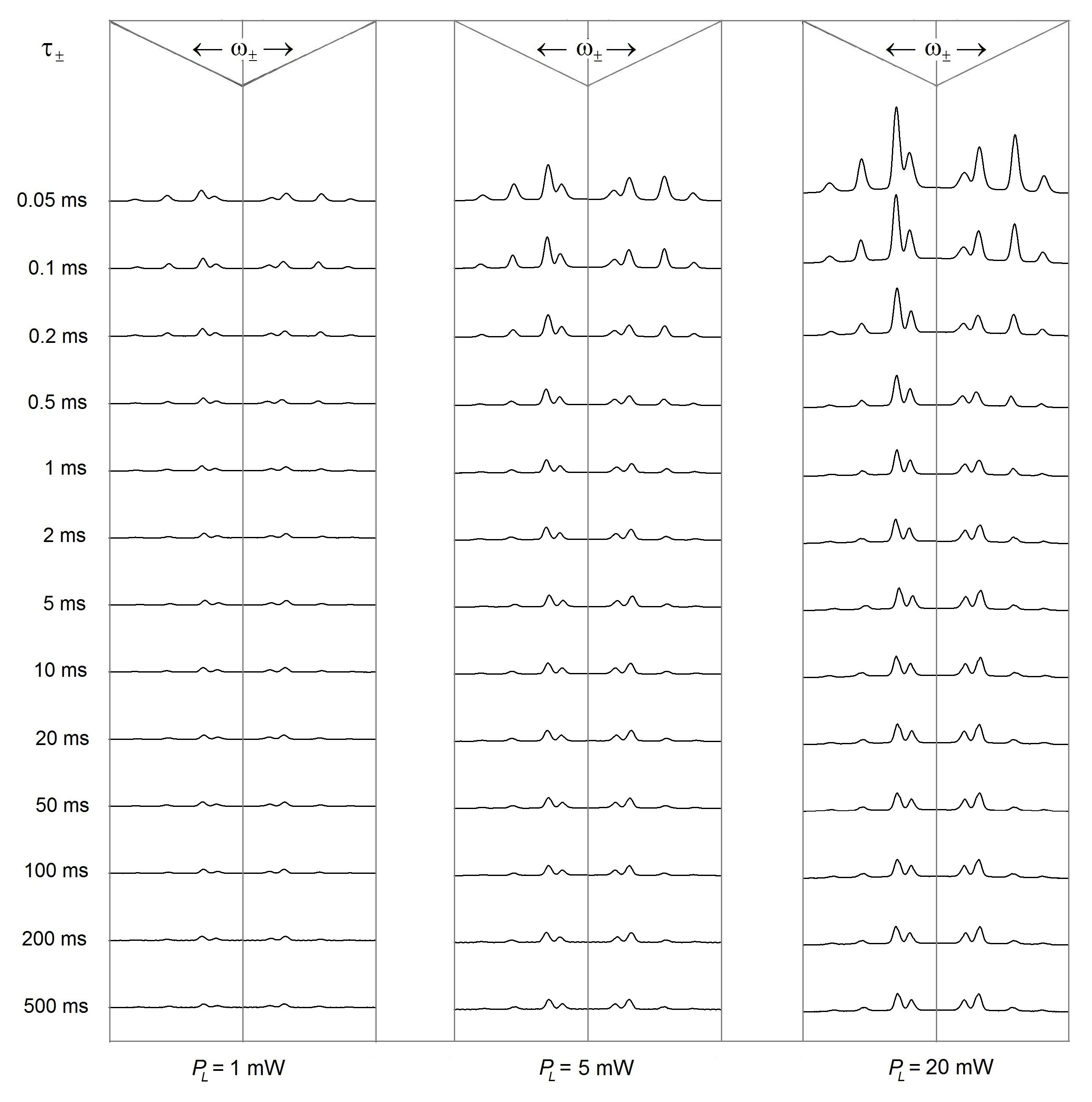}
	\caption{\label{fig:Fig.3} Fluorescence spectra recorded at 13 values of scanning rate for 3 values of laser power: $P_L$ = 1, 5, and 20 mW. $\tau_{\pm}$ indicates rising (right half) and falling (left half) laser frequency scan times. The scanning spectral range is 11.12 GHz, covering $^{87}$Rb F$_g$=2 $\rightarrow$ F$_e$=1,2,3, $^{85}$Rb F$_g$=3 $\rightarrow$ F$_e$=2,3,4, $^{85}$Rb F$_g$=2 $\rightarrow$ F$_e$=1,2,3, and $^{87}$Rb F$_g$=1 $\rightarrow$ F$_e$=0,1,2 transition groups (in order of rising frequency, see right halves of each panel). Spectra are shifted vertically at equal distances; the same vertical scale is applied for all the spectra.}
	\end{center}
\end{figure}

\begin{equation}
	\label{eq::scanrate}
	S=\frac{\partial\omega}{\partial t}=\frac{\Delta\omega_s}{\tau_{\pm}}=2\Delta\omega_sf_s,
\end{equation}

\noindent where $\tau_{\pm}$ is the scanning time on the ascending and descending wings of modulation signal ($\tau_+=\tau_-$). Experimental measurements were done for 13 values of $f_s$, from 1 Hz to 10 kHz (see Table \ref{tab:conversion}). The table contains also corresponding values of $\tau_{\pm}$ and $S/2\pi$.

The recorded spectra are combined in Fig.\ref{fig:Fig.3}. The three column panels represent the results for three values of $P_L$ (1, 5, and 20 mW). In each panel, the spectra recorded for different values of $f_s$ are shifted vertically from each other for visual convenience, preserving a unique vertical scale for the whole graph. First (left) and second (right) halves of the spectrum correspond to falling and rising laser radiation frequency, correspondingly.

\begin{figure*}[ht!]
	\begin{center}
	\includegraphics[width=360pt]{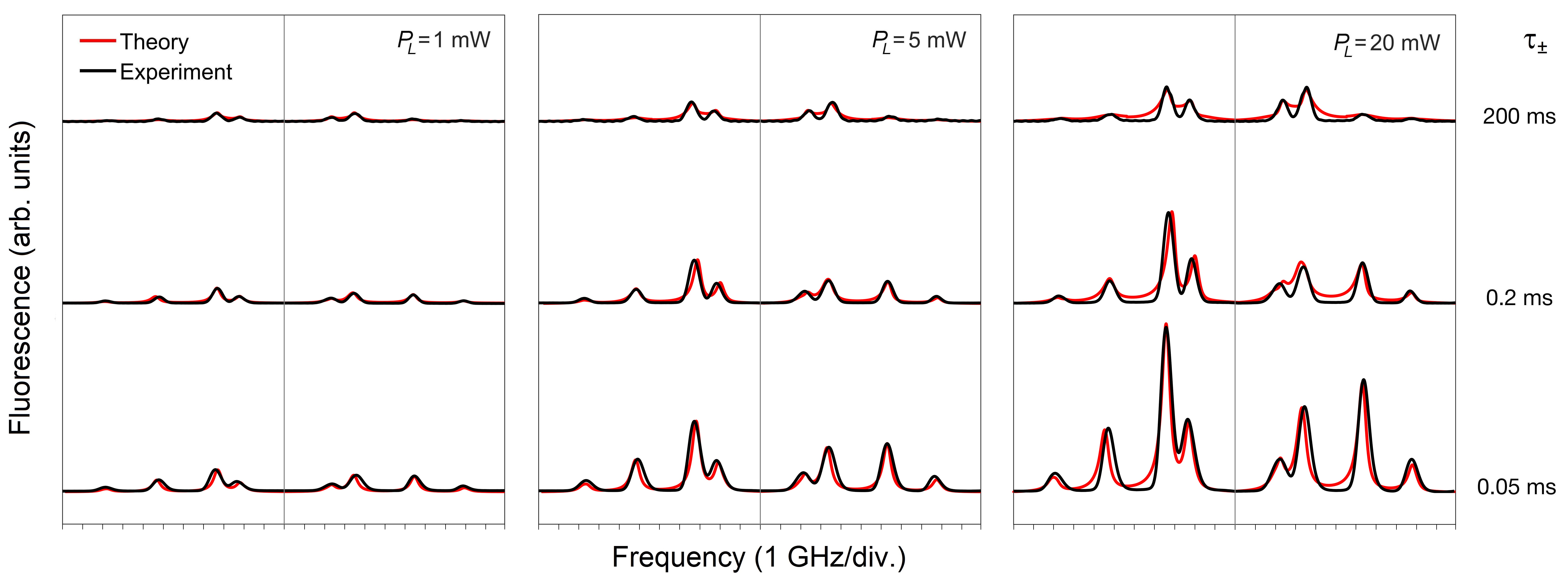}
	\caption{\label{fig:Fig.4} Comparison of theoretical (red lines) and experimental (black lines) fluorescence spectra for 3 values of the laser radiation power and 3 values of the scanning rate.}
	\end{center}
\end{figure*}

The following observations can be drawn from these graphs. For $P_L$ = 1 mW, spectra with $\omega_-$ and $\omega_+$ scans exhibit mirror symmetry, and the shape of spectra does not change significantly when changing $\tau_{\pm}$. Mirror asymmetry in $\omega_-$/$\omega_+$ scans appears for $P_L$ = 5 mW with the decrease of scanning time, over certain values of $\tau_{\pm}$. This asymmetry establishes earlier, and becomes more pronounced for $P_L$ = 20 mW. However, the symmetry tends to recover again when reaching the shortest attainable scan times. Finally, in the slow scanning limit (steady state interaction regime), hyperfine transition groups $^{85}$Rb F$_g$=2 $\rightarrow$ F$_e$=1,2,3 and $^{87}$Rb F$_g$=1 $\rightarrow$ F$_e$=0,1,2 consisting of "open" (non-cycling) components are strongly suppressed, notably for high laser power. Decrease of the scanning time results in gradual enhancement of fluorescence on these transitions.

Numerical simulation of the obtained experimental results using theoretical model described in Section \ref{sec:Sec.2} show a good agreement. Comparison of theoretical and experimental spectra for three values of the scanning rate is presented in Fig.\ref{fig:Fig.4}. The best fitting of spectral lineshapes throughout the whole range of exploited scanning frequencies and incident laser powers has been obtained for $\gamma_0 = 1.03(\pm 0.1)\times 10^{-3} \gamma_{nat}$. 

\begin{figure*}[ht!]
	\begin{center}
	\includegraphics[width=520pt]{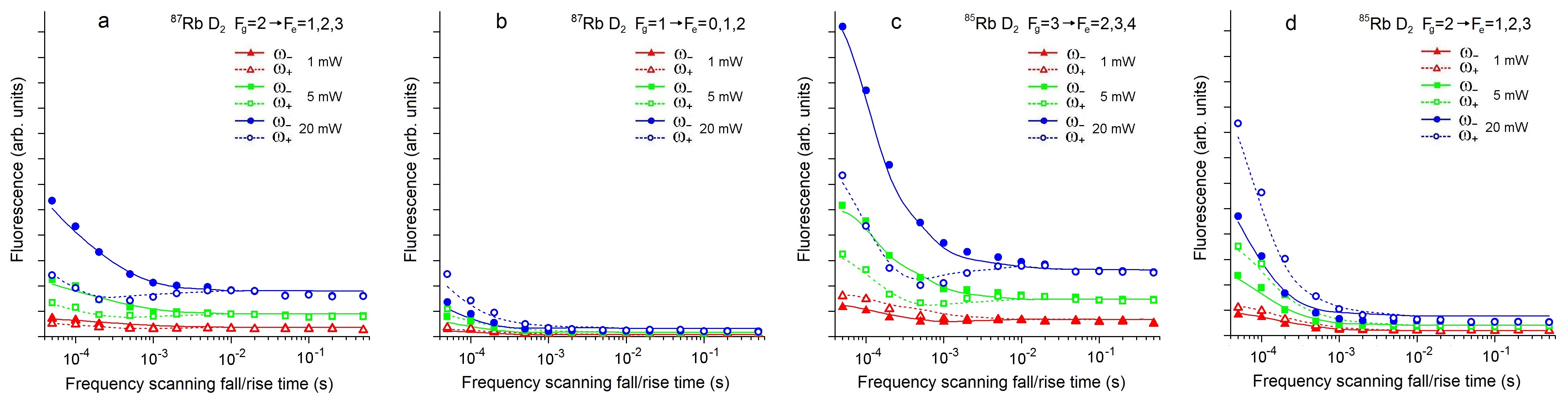}
	\caption{\label{fig:Fig.5} Dependence of the fluorescence peak intensity on the laser frequency scanning rate for the four hyperfine transition groups of Rb D$_2$ line, recorded at three values of $P_L$. Lines: theory; symbols: experiment. Dashed lines / open symbols and solid lines / solid symbols correspond to scanning with falling and rising frequency ($\omega_-$ and $\omega_+$), respectively.}
	\end{center}
\end{figure*}

Quantitative dependences of the fluorescence peak signals from scanning time (separately, for descending $\omega_-$ and ascending $\omega_+$ scans) derived from the spectra shown in Fig.\ref{fig:Fig.3} are presented in Fig.\ref{fig:Fig.5}, along with corresponding theoretical modeling curves. As one can clearly see from these graphs, establishment of steady state interaction regime corresponding to scan-time-independent (horizontal) trace on the graphs, is strongly dependent on laser radiation power, but also somewhat varies for different transition groups. The most drastic changes occur for the groups containing V-type cycling transitions: $^{87}$Rb F$_g$=2 $\rightarrow$ F$_e$=1,2,3 and $^{85}$Rb F$_g$=3 $\rightarrow$ F$_e$=2,3,4, where a deep well is formed at certain value of $\tau_-$ for $\omega_+$ scanning direction, while for the opposite direction $\omega_-$ fluorescence grows monotonically with the decrease of $\tau_+$. Monotonic growth when decreasing scan time is observed also for the transition groups $^{85}$Rb F$_g$=2 $\rightarrow$ F$_e$=1,2,3 and $^{87}$Rb F$_g$=1 $\rightarrow$ F$_e$=0,1,2, independently of the sense of scanning. No expected saturation of this growth was observed at largest values of $f_s$ attainable in our experiment. At the same time, it can be seen that as the scanning speed increases, the peak fluorescence values for scanning with falling and rising frequencies tend to approach each other.

\section{Discussion and emerging results}
\label{sec:Sec.4}

For slow enough scanning the laser radiation frequency allowing establishment of a steady-state atom--light interaction regime, before passing each consecutive resonance with individual atomic transition the atomic system is fully relaxed to equilibrium condition. The situation changes when increasing the scanning rate to a value for which coherence or redistribution of population established during the resonant interaction with a particular hyperfine transition is partly preserved by the time of resonance with the neighbouring transition. This "memory effect" underlies the transient interaction regime, causing modification of magnitudes of individual fluorescence components and its dependence on scanning direction. 

As one can expect, the further increase of scanning rate (beyond the values explored in present experiment) should eventually result in recovery of "linear" response, independent of scanning direction, as no significant redistribution of population can be built because of extremely short interaction time, which is sufficient only for one cycle of absorption and emission. 

The theoretical model used in our work  allows to reproduce experimental results, which indicates that all the involved physical processes are adequately addressed. Throughout the modeling, we have used known spectroscopic parameters for Rb D$_2$ line system, except for two quantities which were free fitting parameters, namely, i) the effective amplitude of the laser electric field $E_0$ dependent on laser radiation power $P_L$, and ii) the relaxation rate of the lower energy levels to the equilibrium isotropic state $\gamma_0$. Necessity to fit the value of $E_0$, which enters in expression for the Rabi frequency $\Omega_{i,j}=\dfrac{\vec{E}_{0} \vec{d}_{i,j}}{\hbar}$, comes from uncertainty of the distribution of laser radiation intensity across the beam and its broad spectral linewidth, which exceeds the Rb natural linewidth.

Much more important fitting parameter is the ground-state relaxation rate $\gamma_0$, which characterizes particular vapor cell used in the experiment. The value of $\gamma_0$ comprises contributions from population relaxation time $T_1$ (relevant for our study) and coherence relaxation time $T_2$. As this parameter remains unchanged throughout the experimental measurements, it can be determined unambiguously by the best fitting of all the experimental spectra recorded for different temporal and power conditions. The obtained fitted value $\gamma_0 \approx 1.03 \times 10^{-3} \gamma_{nat} \approx$ 2$\pi \times$6.25 kHz is close by the order of magnitude to the expected value for the conditions of our experiment. 

Indeed, in the absence of a buffer gas and antirelaxation coating,  the value of relaxation rate $\gamma_0$ should be determined by the flight of optically pumped atoms to the cell walls where they undergo spin-exchange collisions. The contribution from Rb--Rb pairwise collisions has to be completely ruled out because of a small value of the cross-section (1.9$\times$10$^{-14}$ cm$^2$, \cite{Gibbs_1967}) and very low vapor density. As it was shown in \cite{Gharavipour_2017}, the spin-exchange binary collisions in comparable experimental conditions yield only $\sim 2\pi \times$ 25 Hz contribution. In our experiment, the mean atomic velocity $\bar{v}=\sqrt{8k_BT/\pi m}$ = 265 m/s ($k_B$ is Boltzmann constant, $T$ is cell temperature, $m$ is atomic mass). A rough estimate for the atom departing from the laser beam normally towards the wall, undergoing spin-exchange collision with 100 \% probability and returning into the beam with the same trace gives $\gamma_0$ = $2\pi \times$13.3 kHz.

More detailed calculations should take into account the angular distribution of atomic velocity, probabilistic nature of relaxation caused by the atom--wall collision, as well as self-diffusion of atoms in a vapor. All these factors lead to the decrease of the estimated value, as it was demonstrated by Franzen \cite{Franzen_1959}. In the conditions of our experiment, the mean free path of Rb atoms $\lambda=1/\sqrt{2}N\sigma_{self}$, where $N$ is the number density of atoms, and $\sigma_{self}$ is total interatomic collision cross-section, substantially exceeds geometric dimensions of the cell. Indeed, with $N$ = 5$\times$10$^{9}$ cm$^{-3}$ and $\sigma_{self}$ = 1.397$\times$10$^{-13}$ cm$^2$ \cite{Croucher_1968}, we get $\lambda \approx$ 10 m. This estimate indicates that we can consider ballistic trajectories of atoms towards cell walls, excluding as irrelevant the self-diffusion in the vapor.

Elaborated expressions for determination of the ground state population relaxation time $T_1$ (and hence, relaxation rate $\gamma_0$) in a high-vacuum cell, in which interatomic collisions are irrelevant, are presented in \cite{Corney}, where this relaxation time is identified with the mean time of flight of the atoms between two collisions with the walls. Following these calculations, for a cylindric cell with diameter $d$ and length $l$,  the ground state relaxation rate is expressed as:

\begin{equation}
\label{eq::corney}
\gamma_0=2\pi \times \frac{1}{T_1}=2\pi \times \frac{\bar{v}S}{4V}=2\pi \times \frac{\bar{v}(l+d/2)}{ld},
\end{equation}

\noindent where $V$ and $S$ are the cell volume and surface area, respectively. For the conditions of our experiment, Eq.\ref{eq::corney} yields $\gamma_0$ = $2\pi \times$14.2 kHz, which is 2.3 times bigger than the value obtained from the fitting.

This difference can be attributed to the presence of residual buffer gas (unidentified contamination) in the cell, which can lead to the decrease of the estimated value, as it was first shown by Franzen \cite{Franzen_1959}. For a cylindrical cell with diameter $d$ [cm] and length $l$ [cm], the relaxation rate of optically pumped Rb atoms caused by diffusion to the cell walls $\gamma_{0_D}$ [Hz] can be expressed as:

\begin{equation}
	\label{eq::diffus}
	\gamma_{0_D}=2\pi\times \left[ \big( \frac{\mu}{d/2} \big)^2 + \big( \frac{\pi}{l} \big)^2 \right]D
\end{equation}

\begin{equation*}
D=D_0 \frac{p_0}{p} \big(\frac{T}{T_0}\big)^{3/2}
\end{equation*}

\noindent \cite{Gharavipour_2017,Rosenberry_2007}, where $\mu$ = 2.405 is the first zero of the Bessel function, $D$ [cm$^2$/s] is the diffusion coefficient dependent on pressure and temperature, $D_0$ is the diffusion constant at normal conditions (pressure $p_0$ = 760 Torr, temperature $T_0$ = 273 K), $p$ and $T$ are the pressure and temperature of the cell. Taking the values for present work $d$ = 2 cm, $l$ = 13.5 cm, and $\gamma_{0_D} \approx \gamma_0$ = $2 \pi \times$ 6.25 kHz, we obtain for the diffusion coefficient $D \approx$ 1070 cm$^2$/s.

Assuming that our home-made cell was not properly pumped out or that over time (20 years from the date of manufacturing) some air leaked through the welded junctions, we may suppose that the most realistic residual buffer gas is nitrogen (N$_2$). The contribution from known helium permeation through the cell walls is negligible in this case because of low partial pressure in the air ($\sim$ 4 mTorr). As one can find from \cite{Rosenberry_2007}, the diffusion constant for N$_2$--buffered Rb vapor is $D_0$ = 0.144 cm$^2$/s, and from the second expression of Eq.\ref{eq::diffus} for $T$ = 295 K we obtain $p\approx$ 0.11 Torr, which seems realistic.

We should note that if we know the type and pressure of the buffer gas X, this result can be further explored for determination of another important spectroscopic parameter that is a cross-section $\sigma_{buf}$ of elastic velocity-changing Rb--X atomic collisions. The latter can be calculated in the frame of Chapman–Enskog theory, following the expression presented e.g. in \cite{Matushka_2016}:

\begin{equation}
	\label{eq::chapman}
	\sigma_{buf}=\frac{3}{8pD}\sqrt{\frac{\pi(kT)^3}{2m_r}},
\end{equation}

\noindent where $m_r$ is the reduced mass of the interacting particles. Exploring this equation for $p$ = 0.11 Torr, $D$ = 1070 cm$^2$/s, $T$ = 295 K, and reduced mass of Rb--N$_2$ atomic pair $m_r$ = 3.52$\times$10$^{-23}$ g, we obtain the cross-section for Rb--N$_2$ collisions $\sigma_{buf}$ = 4.06$\times$10$^{-15}$ cm$^2$, which is consistent with the value 3.93$\times$10$^{-15}$ cm$^2$ reported in \cite{Croucher_1968}.

The presence of a buffer gas in the cell can be easily checked by implementing the saturated absorption (SA) experiment: addition of > 0.5 Torr of a foreign gas leads to nearly complete suppression of sub-Doppler features in the SA spectrum because of velocity-changing collisions \cite{Hakhumyan_2010}. The SA measurement done with our cell has not revealed any noticeable distinction in appearance of a Doppler pedestal and lineshapes of velocity-selective optical pumping and crossover resonances as compared with a buffer-free reference cell, which indicates that the residual buffer gas pressure is below the critical level. 

We are going to use the results obtained in the present work in our future studies aimed at realization of a heralded control of atomic levels population in alkali metal
vapor by implementing frequency modulation of continuous-wave lasers in a non-stationary
(transient) regime of resonant interaction. The control will be realized by means of changing the
shape, duration, and delay of the sequence of generated pulses. It is expected that the results of these studies can be used for the enhancement of efficiency of photochemical reactions, development of new schemes of sensitive optical magnetometers, development of elements for quantum communication systems, and for other applications. 

\section{Conclusions}
\label{sec:Sec.5}

Summarizing, we have studied the evolution of fluorescence spectra of a room-temperature rubidium vapor in the region of atomic D$_2$ line while changing the linear (triangular) scanning rate of exciting cw laser radiation frequency, exploring changeover from steady-state to transient interaction regime. The general aim of this work was to quantitatively study temporal dynamics of fluorescence basing on extremely simple experiment.

In the low scanning rate limit, the spectral lineshape and magnitude of fluorescence across the hyperfine transitions manifold is independent of the speed and direction of frequency scanning, evidencing the steady-state atom -- radiation field interaction regime. In this regime, the interaction time is determined by the mean time of flight of atoms through the laser beam. Increase of the scanning rate above $\sim$ 2 MHz/$\mu$s (for the conditions of our experiment) results in gradual modification of the amplitudes of fluorescence peaks, different for different transitions and dependent on the scanning direction and speed, manifesting the onset of transient interaction regime. In this regime, the interaction time is caused by the temporal period when the scanned laser field is in resonance with atomic transition. The maximum asymmetry in fluorescence peak amplitudes for rising and falling frequency scanning is obtained at the rate $\sim$ 20 -- 60 MHz/$\mu$s. The symmetry tends to recover again at higher scanning rate.

Theoretical modeling taking into consideration all the relevant physical processes exhibits good agreement with the experimental results. Thanks to this consistency, it is possible to retrieve some important parameters of the experiment, in particular, the relaxation rate of the lower energy levels to the equilibrium isotropic state $\gamma_0$, the diffusion coefficient $D$ in a buffered vapor cell, and the corresponding collisional cross-section $\sigma$.
The obtained results can be also used for determination of appropriate temporal conditions for efficient heralded control of atomic population in a multi-level system, by implementing frequency-modulated cw laser as an effective source of controllable pulsed radiation.

\textbf{Acknowledgments.} We acknowledge the Institute for Informatics and Automation Problems, NAS of Armenia for technical support in numerical calculations. The authors are grateful to D. Sarkisyan and G. Grigoryan for stimulating discussions. This work was supported by the State Committee of Science MES RA, in frame of the research project No.18T-1C234.

\end{document}